\documentclass[12pt]{article}
\usepackage{hhline}
\usepackage{latexsym,graphicx}
\usepackage{subfigure}
\usepackage{amssymb}
\usepackage{amsmath}
\usepackage{amscd}
\usepackage{amsthm}
\usepackage{float}
\usepackage{xcolor}
\usepackage[left=2cm,top=2.5cm,right=2.5cm,bottom=1.5cm]{geometry}    
\begin{document}
\begin{center}
\large{\bf{Cosmological determination to the values of the pre-factors in the logarithmic corrected entropy-area relation}} \\
\vspace{10mm}
\normalsize{Nasr Ahmed$^1$ and Sultan Z. Alamri$^2$}\\
\vspace{5mm}
\small{\footnotesize $^1$ Astronomy Department, National Research Institute of Astronomy and Geophysics, Helwan, Cairo, Egypt\footnote{nasr.ahmed@nriag.sci.eg}} \\
\small{\footnotesize $^2$ Mathematics Department, Faculty of Science, Taibah University, Saudi Arabia\footnote{szamri@taibahu.edu.sa}} \\
\end{center}  
\date{}
\begin{abstract}
In this paper, we continue investigating the possible values of the pre-factors $\alpha$ and $\beta$ in the logarithmic  corrected entropy-area relation based on cosmological stability arguments. In a previous study, we have investigated the stability of the entropy-corrected cosmology using an empirical hyperbolic form of the scale factor. We found that the zero values of the two pre-factors are necessary to obtain a stable flat universe with a deceleration-acceleration transition and no causality violation. The necessity of the zero values of the two pre-factors has also been reached in the current work using a hybrid scale factor Ansatz in the entropy-corrected cosmological equations. Investigating the corrected entropy-area relation in different gravitational and cosmological contexts can provide an accurate estimation to the correct values of the pre-factors. The current work opens a discussion on the validity of the correction terms in the logarithmic corrected entropy-area relation on the cosmological scale. The evolution of the cosmic pressure, energy density, equation of state parameter, jerk parameter and the nonlinear energy conditions has been analyzed.
\end{abstract}

\section{Introduction and motivation}

The discovery of the late-time cosmic acceleration \cite{11,13,14} has been a major challenge to our understanding of gravity and the way it works on cosmological distances. The existence of an unknown energy component with negative pressure (dubbed as dark energy DE) has been assumed to explain this late-time acceleration. Several theoretical models of DE have been constructed through two basic approaches: modifying general relativity \cite{quint}-\cite{nass12} and through scalar fields \cite{1,nass, nass2,moddd,39}, \cite{noj1}-\cite{alt3}.
 A remarkable advance in theoretical physics, which introduced a link between gravity, quantum theory and information, was Bekenstein's suggestion \cite{63} that the black hole's event horizon area is a measure of the black hole's entropy \cite{dav}. Hawking \cite{hawkk} found that black holes emit thermal radiation (Hawking radiation) with a characteristic temperature $T=\frac{\left|\kappa_{sg}\right|}{2\pi}$, where $\kappa_{sg}$ is the surface gravity, and precisely determined the entropy of a black hole as $S=\frac{A}{4G}$. In Hawking radiation, the black hole's entropy $S$ is proportional to its event horizon area $A$. By emitting radiation, the black hole evaporates, its event horizon area decreases, its entropy decreases, but the entropy of the surrounding environment increases due to the emitted radiation. Therefore, a modified version of the second law of thermodynamics (called the generalized second law GSL) has been proposed \cite{63} so that the sum of the entropy of the black hole $S$ and that of the surrounding environment $S_{env}$ cannot decrease, $\frac{d}{dt}(S+S_{env})\geq 0$ . The first law of black hole thermodynamics $T dS = dE$ connects The horizon entropy $S=\frac{A}{4G}$ with the Hawking temperature $T$, where $dE$ is the energy change \cite{62,63,61}. Einstein equations have been derived in \cite{64} using Clausius relation $T dS = \delta Q$ and the horizon- entropy area relation, where $\delta Q$ and $T$ are the energy flux across the horizon and Unruh temperature respectively. A unified first law of black hole dynamics and relativistic thermodynamics $dE=TdS+WdV$ has been derived in spherically symmetric space-times \cite{82}, where $W$ is the work density defined by $-\frac{1}{2} T^{ab}h_{ab}$. Using Clausius relation, Friedmann equations have also been derived from the first law of thermodynamics \cite{65, bo,boo}. It has been indicated that Friedmann equations can be expressed as $dE=TdS+WdV$ at the apparent horizon \cite{60,78} where $E=\rho V$ is the total energy and $W=\frac{1}{2}(\rho-p)$ is the work density. $\rho$ and $p$ are the energy density and pressure of cosmic matter, while $T$ and $S$ are temperature and entropy associated with the apparent horizon. Because the original entropy-area relation is valid only for the case of GR, it needs corrections when considering higher order curvature terms \cite{basic1}. Modified Friedmann equations have been derived in \cite{basic1} by applying the corrected entropy-area relation:
\begin{equation} \label{ent}
S=\frac{A}{4G}+\alpha \ln \frac{A}{4G}+\beta \frac{4G}{A}.
\end{equation}
The pre-factors $\alpha$ and $\beta$ are dimensionless constants whose values are still in debate\cite{salehi}. The correction terms in (\ref{ent}) appear in loop quantum gravity due to quantum fluctuations (see \cite{echde} and references therein). The second correction term has also appeared in the entropic cosmology introduced in \cite{caii}. While positive and negative values have been suggested for $\alpha$ and $\beta$ by some authors \cite{zer1}-\cite{zer5}, it has also been argued that the “best guess” to the logarithmic term might simply be zero \cite{zer}. A stable flat entropy-corrected FRW universe with a deceleration-acceleration transition has been constructed in \cite{nasramri} for zero values of both pre-factors. We believe that studying relation (\ref{ent}) in different gravitational and cosmological contexts can provide an accurate estimation to the correct values of the pre-factors. \par

It is important to discuss the physical/cosmological reasons why the form of the horizon entropy relates to the dynamics of the universe, i.e., the accelerated/decelerated expansion of the universe. Because the 'cosmological event horizon' is the cosmological counterpart of the black hole horizon \cite{counterpart}, it also has an associated entropy proportional to its area \cite{gibbons}. The entropy-area relation remains the same for the cosmological or the black hole horizon. A prominent example is de Sitter universe with its event horizon at a distance $(3/\Lambda)^{\frac{1}{2}}$ from the observer. The entropy of the universe has got a special attention in cyclic cosmology where the cosmological entropy plays a fundamental role in obtaining consistent cyclic models \cite{cy1, cy2, cy3}. While dark energy is widely accepted as a possible explanation for the late-time cosmic acceleration, an alternative entropic explanation has been proposed with the assumption that the apparent cosmological horizon has both a temperature and entropy associated with it \cite{ent4}. Under this assumption, an acceleration term has been obtained in the cosmological equations and a deceleration-acceleration solution has been found.
The entropy-corrected holographic dark energy (ECHDE) \cite{echde} represents another approach to the explain the late-time acceleration based on the corrected entropy-area relation (\ref{ent}), it has the form:
\begin{equation} \label{echde}
\rho_{\Lambda}=3c^2M^2_pL^{-2}+\alpha_1 L^{-4} \ln(M^2_pL^2)+\beta_1 L^{-4},
\end{equation}
Where $M_p$ is the reduced Planck mass, $L$ is the characteristic length scale (the radius of the future event horizon) and $c$ is a constant. While the above ECHDE reduces to the normal holographic dark energy HDE for zero values of $\alpha_1$ and $\beta_1$, the last two terms in (\ref{echde}) becomes effective only at the early stage of the universe when $L$ is very small. \par

The corrected form of the horizon entropy has an important impact on the thermodynamics of the universe. The laws of thermodynamics in a universe with the corrected form of the horizon entropy have been investigated by many authors and in different contexts \cite{therm1, therm2, therm3,therm4, therm5}. In \cite{therm1}, a thermodynamical description of the entropy-corrected holographic dark energy in a universe with a special curvature has been provided using the first law of thermodynamics. Taking into account the quantum corrections to the cosmological horizon entropy, the validity of the generalized second law of thermodynamics (GSL) has been investigated in \cite{therm2} for three different systems. In the same paper, some conditions on the cosmological parameters in the quintessence and phantom eras have been suggested for the validity of GSL. The conditions to validate GSL with corrected entropy in Tachyon Cosmology have been studied in \cite{therm3} with a cosmological dynamical apparent horizon. The validity of GSL with entropy corrections in a flat and closed Kaluza Klein universe has been investigated in \cite{therm4}. The validity of first law and generalized second law of thermodynamics for flat FRW cosmology in Rastall gravity has been investigated analytically in \cite{therm5} for 4 different entropies: 1 - Modified Bekenstein entropy $S_B=\frac{1}{4}(1+\frac{2\gamma}{1+4\gamma})4\pi R_A^2$ where the apparent horizon $R_A$ is given in terms of Hubble parameter as $R_A=\frac{1}{H}$, and is related to Hawking temperature at the apparent horizon $T_A$ by $T_A=\frac{1}{2\pi R_A}$. 2 - Logarithmic corrected entropy given by relation (\ref{ent}). 3 - Power law corrected entropy $S=\frac{\tilde{{A}}}{4l_p^2}(1-F_{\alpha}\tilde{A}^{1-\frac{\alpha}{2}})$, where $F_{\alpha}=\frac{\alpha(4\pi)^{\frac{\alpha}{2}-1}}{(4-\alpha)r_c^{4-\alpha}}$, $r_c$ is the crossover scale and $\alpha$ is a constant. 4 - Renyi entropy $S_R=\frac{\ln(1+\eta S_A)}{\eta}$ where $S_A$ is Tsallis entropy $S_{\delta}=\gamma A^{\delta}$ \cite{therm6}, $\delta$ is the nonadditivity parameter, $\eta$ and $\gamma$ are constants.  \par
While our investigation approach in the current work is based on analyzing the behavior of the cosmic pressure, energy density, equation of state parameter, the jerk parameter and the nonlinear energy conditions, obtaining a viable condition for various dark energy models alternative to the $\Lambda$CDM model has been an active research point. One major requirement for cosmological dark energy solutions is the dynamical stability. A systematic analysis of the dynamical stability of dark energy models has been performed in \cite{indo} for metric-torsion theories. Conditions for the cosmological viability of $f(G)$ modified gravity dark energy models, where $G$ is the Gauss-Bonnet term $G=R^2-4R^{\mu\nu}R_{\mu\nu}+R^{\mu\nu\rho\sigma}R_{\nu\nu\rho\sigma}$, has been found in \cite{indo1} as geometrical constraints on the derivatives. Similar result has been obtained in \cite{indo2} for the cosmological viability of $f(T)$ modified gravity dark energy models, where $T$ is the torsion scalar. Stability and viability conditions for the modified $f(R, T)$ gravity dark energy models, where $R$ is the Ricci scalar and $T$ is the trace of the energy momentum tensor, have been discussed in \cite{indo3}.\newline
Dark energy models and modified gravity theories can also get strongly constrained by observations. In 2017, The gravitational wave observation of the binary neutron star merger GW170817 and the corresponding gamma ray burst (GRB 170817 A) placed tight constraints on the viability of dark energy models constructed in modified gravity theories \cite{indo4, indo5}. The reason is that while most of dark energy theories in modified gravity predict a speed of gravitational waves $c_g$ different from the speed of light $c$, this gravitational wave observation shows that $c_g$ is the same as $c$ within deviations of order $10^{-15}$. The viability of the scalar-tensor theories of gravity with this new information has been discussed in \cite{indo6}. The question if GW170817 can falsify Milgrom's modified Newtonian daynamics (MOND) has been investigated in \cite{indo7}. A review has been given in \cite{indo8} where the surviving dark energy models after GW170817 event has been classified into 4 classes. For the current work based on the corrected entropy-area relation, testing the entropy and thermodynamics of black holes observationally seems impractical task. This is because the flux of Hawking radiation is too small to be distinguished from the surrounding hot environment. Recently, an interesting proposal has been suggested to set a lower limit on the entropy of black holes using gravitational wave observations after the merger of two black holes  \cite{test}. \par
The aim of the current work is to find the best values for $\alpha$ and $\beta$, in the logarithmic corrected entropy-area relation (\ref{ent}), required to describe a stable flat universe in a good agreement with observations \cite{teg,ben,sp}. We use the following hybrid Ansatz which leads to a deceleration-to-acceleration cosmic transition \cite{hyb1}: 
\begin{equation} \label{hyb}
a(t)=a_0\left(\frac{t}{t_0}\right)^{\alpha_1} e^{\beta_1 \left(\frac{t}{t_0}-1\right)},
\end{equation}
where $\alpha$ and $\beta$ are non-negative constants, $a_0$ and $t_0$ are the scale factor and age of the present day universe respectively. Equation (\ref{hyb}) can be reduced after suitable transformations to \cite{kumar}
\begin{equation} \label{hyb2}
a(t)=a_1 t^{\alpha_1} e^{\beta_1 t},
\end{equation}
where $a_1>0$, $\alpha_1 \geq 0$ and $\beta_1 \geq 0$ are constants. This Ansatz is a mixture of power-law and exponential-law cosmologies, and can be regarded as a generalization to each of them. The power-law cosmology can be obtained for $\beta_1=0$, and the exponential-law cosmology can be obtained for $\alpha_1=0$. New cosmologies can be explored for $\alpha_1>0$ and $\beta_1>0$.\par
Since an enormous amount of work has been done in the literature on thermodynamics in the context of cosmology, it is important to illustrate the differences between the current study and the preceding studies. The new ingredients and significant progress of the current work can be summarized in three points. First, The current study investigates the possible values of the pre-factors $\alpha$ and $\beta$ in the logarithmic corrected entropy-area relation based on the stability of the corresponding cosmological solutions. While we consider the best values of the two pre-factors are the values which lead to the most stable solutions, no previous study has concentrated on this cosmological stability approach. We have started this stability approach in \cite{nasramri} via a hyperbolic empirical Ansatz where, surprisingly, the same result of the current work has been obtained: \textit{The most stable cosmological solutions are corresponding to the zero values of the two pre-factors}. Secondly, the suggestion of zero values to both $\alpha$ and $\beta$ (at least on the cosmological scale we are interested in) is new, and it sheds the light on the validity of the correction terms in the logarithmic corrected entropy-area relation (\ref{ent}) on the cosmological scale. A zero value to $\alpha$ in the middle logarithmic term has been suggested in \cite{zer} where it has been shown that it is the unique choice consistent with both the holographic principle and statistical mechanics. Some approaches to the black hole entropy give the values $-\frac{1}{2}$ and $-\frac{1}{3}$ to $\alpha$ in the logarithmic term \cite{log1,log2}. Thirdly, since the values of the pre-factors $\alpha$ and $\beta$ are still in debate, testing the logarithmic corrected entropy-area relation on different scales including the cosmological scale is essential in knowing the relative importance of the last two terms. \par
The paper is organized as follows: The introduction and motivation behind this work is included in section 1. The solution of the cosmological equations with a discussion on the stability and the evolution of different parameters are included in section 2. The final conclusion is included in section 3.

\section{Cosmological equations and solutions} \label{sol}

Considering the corrected entropy-area relation (\ref{ent}), the modified FRW equations can be written as \cite{basic1}
\begin{eqnarray} 
H^2+\frac{k}{a^2}+\frac{\alpha G}{2 \pi}\left(H^2+\frac{k}{a^2}\right)^2-\frac{\beta G^2}{3\pi^2}\left(H^2+\frac{k}{a^2}\right)^3&=&\frac{8\pi G}{3}\rho. \label{cosm1}\\
2\left(\dot{H}-\frac{k}{a^2}\right)\left(1+\frac{\alpha G}{\pi} \left(H^2+\frac{k}{a^2}\right) - \frac{\beta G^2}{\pi^2}\left(H^2+\frac{k}{a^2}\right)^2  \right)&=&-8\pi G (\rho+p).\label{cosm2}
\end{eqnarray}

Where $k=0, 1, -1$ for a flat, closed and open universe respectively. Since recent observations indicate that a cosmic deceleration-to-acceleration transition happened \cite{11,rrr}, new solutions to (\ref{cosm1}) and (\ref{cosm2}) can be explored through empirical forms of the scale factor where the deceleration parameter $q$ changes sign from positive (decelerating universe) to negative (accelerating universe). Taking (\ref{hyb2}) into account, the deceleration parameter $q$ is given as 
\begin{equation} \label{q1}
q(t)=-\frac{\ddot{a}a}{\dot{a}^2}=\frac{\alpha_1}{(\beta_1t+\alpha_1)^2}-1
\end{equation}
The deceleration-to-acceleration transition takes place at $t=\frac{\sqrt{\alpha}-\alpha}{\beta}$ which restricts $\alpha$ in the range $0<\alpha<1$ \cite{kumar}. Solving (\ref{cosm1}) and (\ref{cosm2}) with (\ref{hyb2}), the cosmic pressure $p(t)$ and energy density $\rho(t)$ are expressed as

\begin{eqnarray} \label{p}
p(t)=\frac{1}{16 \pi^3 t^6}\left( \left( 2\beta{\beta_{1}}^{6}-3\pi\alpha{\beta_{1}}^{4}-6{
\pi}^{2}{\beta_{1}}^{2} \right) {t}^{6}-12\beta_{1}\alpha_{1}
 \left( -\beta{\beta_{1}}^{4}+\pi\alpha{\beta_{1}}^{2}+{\pi}^{2}
 \right) {t}^{5}\right. \\  \nonumber
\left. -6\alpha_{1} \left(  \left( -5\beta{\beta_{1}}^{4}+3\pi
\alpha{\beta_{1}}^{2}+{\pi}^{2} \right) \alpha_{1}+\frac{2}{3}\beta{
\beta_{1}}^{4}-\frac{2}{3}\pi\alpha{\beta_{1}}^{2}-\frac{2}{3}{\pi}^{2}
 \right) {t}^{4}\right.~~~~~~~~~~~~~ \\  \nonumber
\left. -12 \left(  \left( -\frac{10}{3}\beta{\beta_{1}}^{2}+\pi\alpha
 \right) \alpha_{1}+\frac{4}{3}\beta{\beta_{1}}^{2}-\frac{2}{3}\pi\alpha
 \right) \beta_{1}{\alpha_{1}}^{2}{t}^{3}\right. ~~~~~~~~~~~~~~~~~~~~~~~~~~~~~\\  \nonumber
\left. -3{\alpha_{1}}^{3} \left(  \left( -10\beta{\beta_{1}}^{2}+\pi
\alpha \right) \alpha_{1}+8\beta{\beta_{1}}^{2}-\frac{4}{3}\pi\alpha
 \right) {t}^{2}+12\beta\beta_{1} {\alpha_{1}}^{4} \left( \alpha_
{1}-\frac{4}{3} \right) t \right. ~~~~~\\  \nonumber
\left. 2\beta {\alpha_{1}}^{5} \left( \alpha_{1}-2 \right)
\right)~~~~~~~~~~~~~~~~~~~~~~~~~~~~~~~~~~~~~~~~~~~~~~~~~~~~~~~~~~~~~~~~~~~~~~~~~~~~~~
\end{eqnarray}
\begin{eqnarray}\label{rho}
\rho(t)=\frac{1}{16\pi^3t^6} \left(\beta_{1}\,t+\alpha_{1} \right) ^{2}\left( (3\pi \alpha \beta_1^2-2\beta \beta_1^4+6\pi^2)t^4\right.~~~~~~~~~~~~~~~~ \\  \nonumber
\left. \alpha_1\beta_1(6\pi\alpha-8\beta)t^3+\alpha_1^2(3\pi\alpha-12\beta\beta_1^2)t^2-8\beta\beta_1\alpha_1^3t-2\beta\alpha_1^4\right)
\end{eqnarray}
The EoS parameter $\omega(t)=\frac{p(t)}{\rho(t)}$ is simply the division of (\ref{p}) and (\ref{rho}). Determining the value of the EoS parameter is essential to investigate the nature of dark energy. The value of this parameter is $0$ for dust, $1/3$ for radiation and $-1$ for the current cosmological constant (dark energy) epoch. $\omega \leq -1$ for phantom scalar field and $-1 \leq \omega \leq 1$ for quintessence scalar field. It can also evolve across the phantom divide line $\omega = -1$ for quintom field. $\omega=1$ is the largest value of this parameter consistent with causality and is supposed to happen for some exotic type of matter called stiff matter \cite{zeld} where the sound speed equals the speed of light.
The jerk parameter is defined as \cite{j1,j2} 
\begin{equation}\label{jerk}
j=\frac{\dddot{a}}{aH^3}=q+2q^2-\frac{\dot{q}}{H}
\end{equation}
where $q$ is the deceleration parameter. Since $j = 1$ for flat $\Lambda CDM$ models \cite{j3}, this parameter helps to describe models close to $\Lambda CDM$. The value of $j$ for the current model is
\begin{equation}
j=\frac{\beta_1 t+\alpha_1}{t}
\end{equation}
Fig. 1(a) shows that the deceleration parameter varies in the range $-1\leq q \leq 1$. It starts at $q=1$ (a decelerating radiation-dominated era), crosses the decelerating matter-dominated era at $q=\frac{1}{2}$, changes sign to negative (accelerating era) and ends at $q=-1$ (de Sitter universe). The evolution of the jerk parameter shows that it tends to $1$ at late-times where the current model becomes in a good agreement with the flat $\Lambda CDM$ model. The cosmic pressure $p$ also changes sign from positive in early decelerating time where attractive gravity dominates, to negative in late accelerating time where repulsive gravity (represented in dark energy) dominates. We have tried all possible values of the four basic parameters in the current model, namely $\alpha$, $\beta$, $\alpha_1$ and $\beta_1$. The possibility of a causality violation where the EoS parameter exceeds unity ($\omega(t) >1$) exists for all values of $\alpha$ and $\beta$ except when both of them are zero where we obtain $-1 \leq \omega(t) \lesssim \frac{1}{3}$. Consequently, the evolution of the EoS parameter in the current model strongly supports the zero values of the pre-factors. It has also been shown in \cite{zer} that the zero value of $\alpha$ is the unique choice consistent with both the holographic principle and statistical mechanics. Table (1) shows the behaviour of the cosmic pressure, energy density, EoS parameter and the new nonlinear energy conditions for different positive, negative and zero values of $\alpha$, $\beta$, $\alpha_1$ and $\beta_1$. We have found that some choices of $\alpha$ and $\beta$ are not allowed where the energy density $\rho(t)$ shows a wrong behavior and goes to $-\infty$ as $t \rightarrow 0$. We can also see from the table that the most stable and physically acceptable solution happens when the pre-factors $\alpha$ and $\beta$ take zero values. The evolution of the EoS parameter shows no quintom behavior (no cosmological constant boundary crossing) as the lower bound is $-1$ for all possible values of $\alpha$ and $\beta$. \par

Because the correct values of the pre-factors in the corrected entropy-area relation (\ref{ent}) are still in debate, studying this relation in different contexts, gravity theories and different setups is very helpful in determining the correct values. It is interesting to note that the same result we have obtained here on the zero values of $\alpha$ and $\beta$ using the hybrid law, has also been reached using the hyperbolic law $a(t)=A\, (\sin(\zeta t))^{\frac{1}{2}}$ when solving the cosmological equations (\ref{cosm1}) and (\ref{cosm2}) \cite{nasramri}. As we have indicated in \cite{nasramri}, the hyperbolic scale factor $a(t)=A\, (\sin(\zeta t))^{\frac{1}{2}}$ also allows a deceleration-to-acceleration transition, its jerk parameter tends to a flat $\Lambda CDM$ ($j=1$) at late-times, and it appears in many contexts of cosmology (see \cite{nasramri} and references therein). Fig.1(c) shows that pressure is positive during the early-time decelerating era and negative during the late-time accelerating era. Because of the presence of semiclassical quantum effects in the current model, we have tested the physical acceptability of the model through the new nonlinear energy conditions \cite{ec,FEC1, FEC2, detec, nasramri} which are: (i) The flux energy condition (FEC): $\rho^2 \geq p_i^2$ \cite{FEC1, FEC2}. (ii) The determinant energy condition (DETEC): $ \rho\, .\, \Pi p_i \geq 0$ \cite{detec}. (ii) The trace-of-square energy condition (TOSEC): $\rho^2 + \sum p_i^2 \geq 0$ \cite{detec}. All of them are satisfied for the current entropy-corrected model. Table1 shows that (1) The possibility of a causality violation exists for all values of $\alpha$ and $\beta$ except for $\alpha=\beta=0$ where $-1 \leq \omega(t) \lesssim \frac{1}{3}$. (2) The most stable solution is obtained for the flat universe ($k=0$) with zero values of the pre-factors $\alpha$ and $\beta$. 

\section{Conclusion}

The possible values of the pre-factors in the logarithmic corrected entropy-area relation have been investigated based on cosmological stability. The main results are:  \par
\begin{itemize}
  \item The best values for $\alpha$ and $\beta$ required to describe a stable flat universe with a deceleration-to-acceleration transition and no causality violation are the zero values.
  \item The cosmic pressure is positive during the early-time decelerating epoch and negative during the late-time accelerating epoch. 
	 \item The violation of causality is possible for all values of $\alpha$ and $\beta$ except for $\alpha=\beta=0$ where $-1 \leq \omega(t) \lesssim \frac{1}{3}$.
\end{itemize}
Same results have also been obtained in a previous study using the empirical hyperbolic law $a(t)=A\, (\sin(\zeta t))^{\frac{1}{2}}$ in \cite{nasramri}. Predicting zero values of $\alpha$ and $\beta$ in two different cosmological solutions represents a strong support for the zero values of the two pre-factors on the cosmological scale. 

\begin{table}[H]\label{tap}
\centering
\tiny
    \begin{tabular}{ | p{1.8cm} | p{2cm} | p{2cm} | p{2cm} | p{2cm} | p{2cm} | p{2.2cm} | }
    \hline
     $\alpha$   & 0   & 0   & 0.2&0.2, ~0.3,~0.4,~0.5  &0&0\\ \hline
    $\beta$     & 0   & 0.2, ~0.3,~0.4,~0.5 & 0.2& 0& 0&0\\ \hline
		 $\alpha_1$ & 0.5 & 0.5 & 0.5 & 0.5& 0.1&0.4 \\ \hline
		$\beta_1$   &0.5  & 0.5 &0.5 & 0.5& 0.1&0.4 \\ \hline
    $\rho \rightarrow \infty$ as $t \rightarrow 0$ & $\checkmark$ & $\times$ & $\times$  &$\checkmark$  & $\checkmark$ & $\checkmark$  \\ \hline
	   $p$:(+ve)$ \rightarrow$(-ve)                  & $\checkmark$ & (-ve) &(-ve)  & $\checkmark$ & (+ve) &$\checkmark$  \\ \hline
		 $\omega(t)$ & $-1 \leq \omega(t) \leq \frac{1}{3}$ &  $-1 \leq \omega(t) \leq 3$ & $\rho \rightarrow -\infty$ as $t \rightarrow 0$   &  $-1 \leq \omega(t) \lesssim 1.7$ & $-1 \leq \omega(t) \lesssim 5.7$  & $-1 \leq \omega(t) \lesssim 0.7$ \\ \hline
		 FEC & $\checkmark$ & $\rho \rightarrow -\infty$ as $t \rightarrow 0$ & $\rho \rightarrow -\infty$ as $t \rightarrow 0$  & $\checkmark$ &  $\times$& $\checkmark$ \\ \hline
		DEC & $\checkmark$ & $\rho \rightarrow -\infty$ as $t \rightarrow 0$ &  $\rho \rightarrow -\infty$ as $t \rightarrow 0$  & $\checkmark$ & $\checkmark$ & $\checkmark$  \\ \hline
		TSEC & $\checkmark$ & $\rho \rightarrow -\infty$ as $t \rightarrow 0$&  $\rho \rightarrow -\infty$ as $t \rightarrow 0$  & $\checkmark$ & $\checkmark$ & $\checkmark$  \\ \hhline{|=|=|=|=|=|=|=|}
		  $\alpha$ & 0 & 0 & 0& 0&0 &0 \\ \hline
    $\beta$    & 0 & 0 & 0& 0& 0&0\\ \hline
		$\alpha_1$ & 0.6 & 0.3& $\frac{1}{3}$&  $\frac{1}{4}$& $\frac{1}{3}$&2 \\ \hline
		$\beta_1$ & 0.6 & 0.3 & $\frac{1}{3}$&  $\frac{1}{4}$& 2&$\frac{1}{3}$ \\ \hline
    $\rho \rightarrow \infty$ as $t \rightarrow 0$ & $\checkmark$ &  $\checkmark$ & $\checkmark$  & $\checkmark$& $\checkmark$&$\checkmark$  \\ \hline
	   $p$:(+ve)$ \rightarrow$(-ve) & $\checkmark$ &  $\checkmark$ &$\checkmark$  & $\checkmark$ & $\checkmark$ & (-ve) \\ \hline
		 $\omega(t)$ & $-1 \leq \omega(t) \leq 0.1$ & $-1 \leq \omega(t) \leq 1.2$ & $-1 \leq \omega(t) \leq 1$  & $-1 \leq \omega(t) \lesssim 1.7$  &  $-1 \leq \omega(t) \lesssim 1.7$ & $-1 \leq \omega(t) \lesssim -0.67$  \\ \hline
		 FEC & $\checkmark$  & $\checkmark$ & $\checkmark$   & $\checkmark$ at late-time & $\checkmark$ & $\checkmark$ \\ \hline
		DEC & $\checkmark$ & $\checkmark$ &$\checkmark$  & $\checkmark$ & $\checkmark$ & $\times$ \\ \hline
		TSEC & $\checkmark$  & $\checkmark$ & $\checkmark$  & $\checkmark$ & $\checkmark$ & $\checkmark$ \\ \hhline{|=|=|=|=|=|=|=|}
    \end{tabular}
		\caption {The behavior of $p(t)$, $\rho(t)$, $\omega(t)$ and the nonlinear energy conditions for different values of $\alpha$, $\beta$, $\alpha_1$ and $\beta_1$.}
		\end{table}

\begin{figure}[H]
  \centering            
  \subfigure[$q$]{\label{F63}\includegraphics[width=0.2\textwidth]{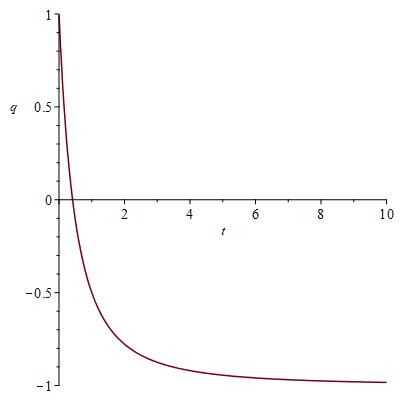}} 
		 \subfigure[$j$]{\label{F6006}\includegraphics[width=0.2\textwidth]{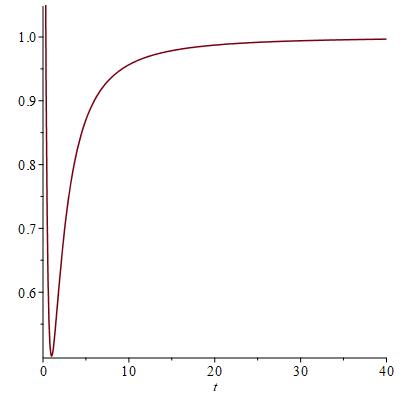}}
	 \subfigure[$p$]{\label{F423}\includegraphics[width=0.2\textwidth]{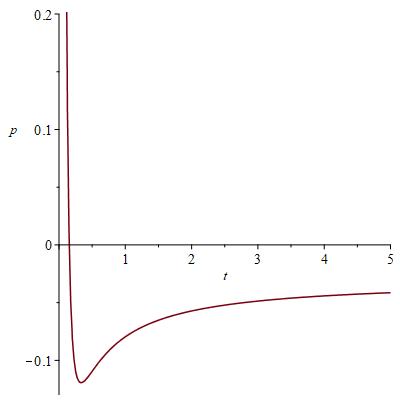}}  
	\subfigure[$\rho$]{\label{F4273}\includegraphics[width=0.2\textwidth]{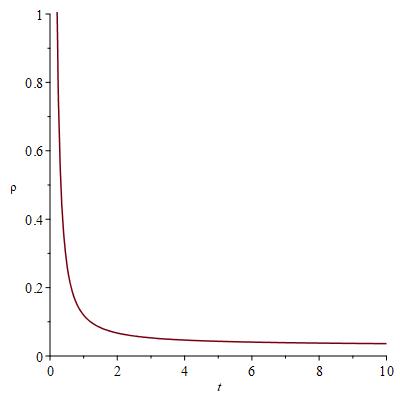}}\\ 
	\subfigure[$\omega$]{\label{F4}\includegraphics[width=0.2\textwidth]{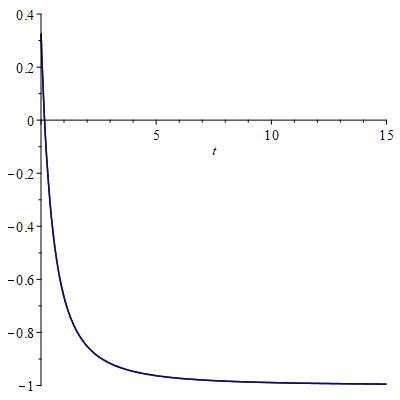}}
	  \subfigure[$\rho^2-p^2$]{\label{F636}\includegraphics[width=0.2\textwidth]{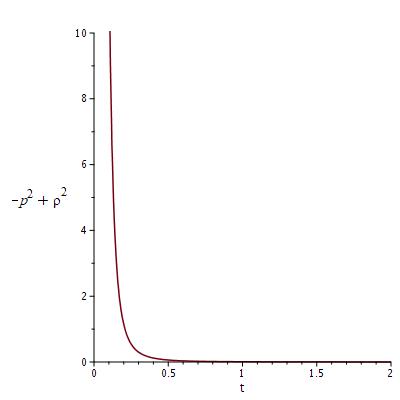}} 
		\subfigure[$\rho . p^3$]{\label{F649}\includegraphics[width=0.2\textwidth]{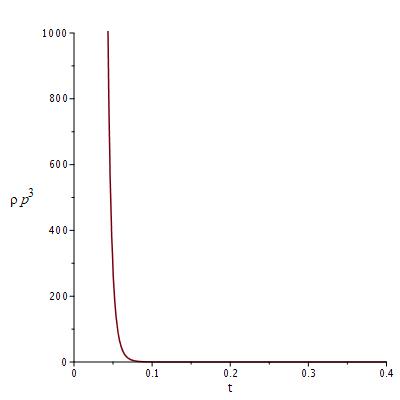}} 
	\subfigure[$\rho^2 . 3p^2$]{\label{F6498}\includegraphics[width=0.2\textwidth]{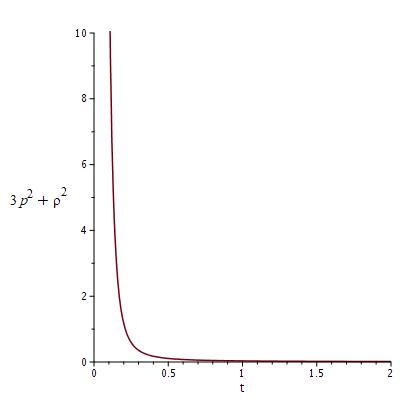}}
  \caption{Fig. 1(a) The deceleration parameter varies in the range $-1\leq q \leq 1$. (b) The jerk parameter $j=1$ at late-times where the current model tends to a flat $\Lambda CDM$ model. (c) The cosmic pressure shows a sign flipping from positive to negative. (d) the energy density is always positive. (e) $\omega(t)$ varies in the range $-1\leq q \leq \frac{1}{3}$. (f), (g) and (H) show the validity of the three nonlinear energy conditions. Here $\alpha=\beta=0$ and $\alpha_1=\beta_1=0.5$.}
  \label{fig:cassimir55}
\end{figure}


\begin{thebibliography}{paper}
\bibitem{11} S. Perlmutter et al., Astrophys. J. 517, 565 (1999).
\bibitem{rrr} A. Riess et al., Astron. J. 116, 1009 (1998).
\bibitem{13} W. J. Percival et al., Mon. Not. Roy. Astron. Soc. 327, 1297 (2001).
\bibitem{14} D. Stern, R. Jimenez, L. Verde, M. Kamionkowski \& S. A. Stanford, J. Cosm. Astrop. Phys. 008, (2010).
\bibitem{quint} S. Tsujikawa, Class. Quant. Grav. 30, 214003 (2013).
\bibitem{chap} A. Y. Kamenshchik, U. Moschella \& V. Pasquier Phys. Lett. B511, 265 (2001).
\bibitem{phant} R. R. Caldwell, Phys. Lett. B 545, 23 (2002).
\bibitem{ess} T. Chiba, T. Okabe \& M. Yamaguchi, Phys. Rev. D62, 023511 (2000).
\bibitem{tak} A. Sen, JHEP 0207, 065 (2002). 
\bibitem{ark} N. Arkani-Hamed, H. Cheng, M. A. Luty \& S. Mukohyama, JHEP 0405, 074 (2004).
\bibitem{nass1} Nasr Ahmed \& Sultan Z. Alamri, Res. Astron. Astrophys., Vol 18, No. 10 (2018). 
\bibitem{nass11} Nasr Ahmed \& Sultan Z. Alamri, Can. J. Phys. (2019) in press https://doi.org/10.1139/cjp-2018-0635.
\bibitem{nass12} Nasr Ahmed, arXiv:1901.04849 [gr-qc].
\bibitem{1} T. Harko, F. S. N. Lobo, S. Nojiri \& S. D. Odintsov, Phys. Rev. D 84, 024020 (2011).
\bibitem{nass} Nasr Ahmed \& I. G. Moss, JHEP 12, 108 (2008).
\bibitem{nass2} Nasr Ahmed \& I. G. Moss, Nuclear Physics B 833,1-2 (2010).
\bibitem{moddd} S. Nojiri,  S. D. Odintsov \& V. K. Oikonomou, Phys. Rept. 692 (2017).
\bibitem{39} S. Nojiri \& S. D. Odintsov, Phys. Rev. D 74, 086005 (2006).
\bibitem{noj1} S. Nojiri \& s. D. Odintsov, Phys. Rev. D74, 086005 (2006).
\bibitem{noj8} S. Nojiri, S. D. Odintsov \& P. V. Tretyakov, Prog. Theor. Phys. Suppl. 172, 81 (2008).
\bibitem{torsion} R. Ferraro \& F. Fiorini, Phys. Rev. D 75, 084031 (2007).
\bibitem{beng1} G. R. Bengochea \& R. Ferraro, Phys. Rev. D 79, 124019 (2009).
\bibitem{de1} A. De Felice \& S. Tsujikawa, Living Rev. Rel. 13, 3 (2010).
\bibitem{alt1} M. E. S. Alves, O. D. Miranda \& J. C. N. de Araujo, 700 (5), (2011).
\bibitem{alt2} A. Maeder, The Astro. J. Vol 849, 2 (2017).
\bibitem{alt3} J. Gagnon and J. Lesgourgues, J. Cosmol. Astropart. Phys. 09 (026), 2011 [1107.1503]
\bibitem{63} J. D. Bekenstein, Phys. Rev. D7, 2333 (1973).
\bibitem{dav} T. M. Davies, P. C. W. Davies \& C. H. Lineweaver, Class. Quant. Grav. 20, 2753-2764 (2003).
\bibitem{hawkk} S. W. Hawking, Comm. Math. Phys. 25, 152–166 (1972), Commun. Math. Phys. 43, 199 (1975).
\bibitem{61} j. M. Bardeen, B. Carter \& S. W. Hawking, Commun. Math. Phys.
31, 161 (1973).
\bibitem{62} S. W. Hawking, Commun. Math. Phys. 43, 199 (1975).
\bibitem{64} T. Jacobson, Phys. Rev. Lett. 75, 1260 (1995).
\bibitem{82} S. A. Hayward, S. Mukohyana \& M. C. Ashworth, Phys. Lett. A256, 347(1999).
\bibitem{65} R. G. Cai \& S. P. Kim, JHEP 02, 050 (2005).
\bibitem{bo} R. Bousso, JHEP. 9907, 004 (1999).
\bibitem{boo} S. Nojiri \& S. D. Odintsov, Gen. Relativ. Grav. 38, (2006) 1285.
\bibitem{60} M. Akbar \& R. G. Cai, Phys. Rev. D75, 084003 (2007).
\bibitem{78} S. A. Hayward, S. Mukohyana \& M. C. Ashworth, Phys. Lett. A 256, 347 (1999).
\bibitem{basic1} R. G. Cai, L. Cao, H. Ya-Peng, JHEP 0808, 090 (2008) .
\bibitem{salehi} A. Salehi \& M. Fard, Eur. Phys. J. C 78:232 (2018).
\bibitem{echde} H. Wei, Commun. Theor. Phys. 52, 743 (2009).
\bibitem{caii} Y. F. Cai, J. Liu \& H. Li, Phys. Lett. B 690, 213, (2010).
\bibitem{zer1} J. L. Jing and M. L. Yan, Phys. Rev. D 63, 024003 (2001).
\bibitem{zer2} G. Gour, Phys. Rev. D 66, 104022 (2002).
\bibitem{zer3} S. Hod, Higher-order corrections to the entropy and area of quantum black holes, arXiv:hep-th/0405235 (2004).
\bibitem{zer4} J.Q. Xia, H. Li, X. Zhang, Phys. Rev. D 88, 063501 (2013).
\bibitem{zer5} W. Yang, L. Xu, Phys. Rev. D 89,  083517 (2014).
\bibitem{zer} A. J. M. Medved, Class. Quant. Grav. 22, 133 (2005).
\bibitem{nasramri} Nasr Ahmed and Sultan Z. Alamri, arXiv:1811.08864 [gr-qc].
\bibitem{counterpart} T. K. Mathew, R. A. \& V. K. Soman, Eur. Phy. J. C. 73 (2013). 
\bibitem{gibbons} G. W. Gibbons \& S. W. Hawking, Phys. Rev. D 15, 2738 (1977).
\bibitem{cy1}C.A. Egan and C.H. Lineweaver, Astrophys. J. 710, 1825 (2010). arXiv:0909.3983.
\bibitem{cy2} P.H. Frampton, JCAP 0910, 016 (2009). arXiv:0905.3632 [hep-th].
\bibitem{cy3}	Nasr Ahmed \& Sultan Z. Alamri, Canadian Journal of physics (2019) in press.  https://doi.org/10.1139/cjp-2018-0635
\bibitem{ent4} A. Damien et. al. Phys. Lett. B. 696 (3) (2011).
\bibitem{therm1} M. Jamil, A. Sheykhi \& M. Farooq, Int. J. Mod. Phys. D 19:1831-1842 (2010).
\bibitem{therm2} M. Sharif \& Abdul Jawad, Int. J. Mod. Phys. D. 22 (3)1350014 (2013).
\bibitem{therm3} H. Farajollahi, Astrophys. Space Sci. 350, 325-331 (2014).
\bibitem{therm4} M. Sharif \& Abdul Jawad, Chin. J. Phys. 35, 6 (2015).
\bibitem{therm5} K. Bamba et. al., Eur. Phys. J. C 78, 986 (2018).
\bibitem{therm6} C. Tsallis \& L. J. L. Cirto, Eur. Phys. J. C 73, 7  (2013).
\bibitem{indo} A. S. Bhatia \& S. Sur, Int. J. Mod. Phys. D 26 (2017) 1750149.
\bibitem{indo1} S. Zhou, E. J. Copeland \& Paul M. Saffin, JCAP V 2009 (2009). 
\bibitem{indo2} M.R. Setare \& N. Mohammadipour, JCAP, V 2012 (2012).
\bibitem{indo3} E. H. Baffou et. al., Astrophys. Space Sc.356, 1 (2015).
\bibitem{indo4} Virgo, LIGO Scientific Collaboration, B. Abbott et. al., Phys. Rev. Lett. 119 (2017), no. 16 161101, 1710.05832.
\bibitem{indo5} Virgo, Fermi-GBM, INTEGRAL, LIGO Scientific Collaboration, B. P. Abbott et. al., Astrophys. J. 848 (2017), no. 2 L13, 1710.05834.
\bibitem{indo6} D. Langlois et. al., Phys. Rev. D 97, 061501 (2018).
\bibitem{indo7} R.H. Sanders, Int. J. Mod. Phys. D 27, 14 (2018).
\bibitem{indo8} R. Kase \& S. Tsujikawa,  Int. J. Mod. Phys. D 28, 5 (2019).
\bibitem{test} R. Brustein, A.J.M. Medved \& K. Yagi, CERN-TH-2018-261.
\bibitem{teg} M. Tegmark et al., Phys. Rev. D 69, 103501 (2004).
\bibitem{ben} C.L. Bennett et al., Astrophys. J. Suppl. 148, 1 (2003).
\bibitem{sp} D.N. Spergel et al., Astrophys. J. Suppl. 148, 175 (2003a).
\bibitem{hyb1} Özgür Akarsua et. al. JCAP 2014 (2014) arXiv:1307.4911.
\bibitem{log1} R.K. Kaul, P. Majumdar, Phys. Rev. Lett. 84, 5255 (2000).
\bibitem{log2} M. Domagala, J. Lewandowski, Class. Quant. Grav. 21, 5233 (2004).
\bibitem{kumar} S. Kumar, Gravit. Cosmol. 19 : 284 (2013).
\bibitem{zeld} Y. B. Zeldovich, MNRAS, 160, 1P ( 1972).
\bibitem{j1}T. Chiba and T. Nakamura, Prog. Theor. Phys. 100, 1077 (1998).
\bibitem{j2} M. Visser, Class. Quantum Grav. 21, 2603 (2004); M. Visser, Gen. Relativ. Gravit. 37, 1541 (2005).
\bibitem{j3} D. Rapetti, S. W. Allen, M. A. Amin, and R. D. Blandford, astro-ph/0605683
(2006).
\bibitem{ec} P. Martın–Moruno and M. Visser, JHEP 1309, 050 (2013) .
\bibitem{FEC1} G. Abreu, C. Barcel´o \& M. Visser, J.High Energy Physics JHEP12, 092 (2011).
\bibitem{FEC2} P. Mart´ın-Moruno \& M. Visser, Phys. Rev. D 88 (6) 061701 (2013).
\bibitem{detec} P. MartnMoruno and M. Visser, JHEP 1309, 050 (2013).


\end{thebibliography}
\end{document}